# Lowering Insulator-to-Metal Transition Temperature of Vanadium Dioxide Thin Films via Co-Sputtering, Furnace Oxidation and Thermal Annealing

Vishwa Krishna Rajan,[#] Jeremy Chao,[#] Sydney Taylor, and Liping Wang[*]

*School for Engineering of Matter, Energy and Transport*

*Arizona State University, Tempe, Arizona 85287*

[#] Equal contributions

[*] Corresponding author. liping.wang@asu.edu

**ABSTRACT**

Thermochromic vanadium dioxide thin films have attracted much attention recently for constructing variable-emittance coatings upon its insulator-metal phase transition for dynamic thermal control. However, fabrication of high-quality vanadium dioxide thin films in a cost-effective way is still a challenge. In addition, the phase transition temperature of vanadium dioxide is around 68°C, which is higher than most of terrestrial and extraterrestrial applications. In this study, we report the fabrication and characterization of tungsten-doped vanadium dioxide thin films with lowered phase transition temperatures via co-sputtering, furnace oxidation and thermal annealing processes for wider application needs. The doping is achieved by co-sputtering of tungsten and vanadium targets while the doping level is varied by carefully controlling the sputtering power for tungsten. Doped thin film samples of 30-nm thick with different tungsten atomic concentrations are prepared by co-sputtering onto undoped silicon wafers. Optimal oxidation time of 4 hours is determined to reach full oxidation in an oxygen-rich furnace environment at 300°C. Systematic thermal annealing study is carried out to find the optimal annealing temperature and time. By using an optical cryostat coupled to an infrared spectrometer, the temperature-dependent infrared transmittance of fully annealed tungsten-doped vanadium dioxide thin films are measured in a wide temperature range from -60°C to 100°C. The phase transition temperature is found to decrease at 24.5°C per at.% of tungsten doping, and the thermal hysteresis between heating and cooling shrinks at 5.5°C per at.% from the fabricated vanadium dioxide thin films with tungsten doping up to 4.1 at.%.

# I. INTRODUCTION

Vanadium dioxide ($VO_2$) is a thermochromic material with a reversible insulator-to-metal phase transition (IMT) around its transition temperature of 68°C.[1-3] The phase change between monoclinic insulator and rutile metal is not instantaneous and depends on the fabrication method of the vanadium dioxide as well as the heating and cooling rate.[4-9] This thermochromic property allows vanadium dioxide to be utilized in many radiative heat transfer applications where passive thermal control is desired. Thermochromic smart windows regulate the incoming solar radiation in order to minimize radiative heat transfer when temperatures are high while allowing this heat to transmit through the window when temperatures are low, allowing for temperature control of an indoor space without additional power requirements.[7,10-13] Modulating the infrared transmittance through a thermochromically switchable filter can also be used in other applications, such as thermal camouflaging or energy dissipation techniques.[14,15]. Another popular use of vanadium dioxide is in infrared emitters, where the emissivity of the device is high at high temperatures, promoting heat rejection away from the device and into the space, but at low temperatures the emissivity is low to prevent heat loss.[16] These devices are accomplished by incorporating vanadium dioxide into Fabry-Perot metafilm designs[17-21] to take advantage of resonance effects of the metallic vanadium dioxide in the design, or using vanadium dioxide based metamaterials[22-27] where nanostructures are used to excite different effects depending on the phase of the vanadium dioxide. The deposition of vanadium dioxide on flexible substrates has also been a topic of recent interest and allows for vanadium-dioxide-based devices to find even broader application.[28,29]

However, one challenge of these devices is the relatively high transition temperature of vanadium dioxide. For many terrestrial thermal control applications, the desired transition temperature should be at or near room temperature to promote a comfortable thermal environment for humans. For extraterrestrial spacecraft applications, the desired transition temperature may be even lower depending on the surrounding temperature of a particular space mission. Additionally, the thermal hysteresis of the phase transition should be narrow, such that the device can quickly switch between its high and low emittance/transmittance states. While the transition temperature and phase hysteresis of vanadium dioxide are affected by other factors such as fabrication method, and thin-film layer parameters such as thickness, crystal structure, mechanical stress or deformation, and non-stoichiometry, doping the vanadium dioxide with other elements has been a popular and effective method in lowering the transition temperature.

Substitutional dopant elements, which could alter the crystalline and electronic structures of $VO_2$, can be used to lower the transition temperature.[30-34] Tungsten dopants can reduce the transition temperature by 20-30°C per at.% of tungsten,[30,33,34] indicating that around 2 at.% of tungsten would be effective in lowering the transition temperature from 68°C to room temperature. There are quite a few fabrication methods that can be used to introduce tungsten dopants into vanadium dioxide thin films, among them physical vapor deposition and wet chemistry methods are two popular fabrication methods. Physical vapor deposition involves a solid vanadium source material being converted to a vapor, transferred to a substrate, then condensed back as a thin film. The thin film on the substrate could be a precursor vanadium film to be oxidized with a later process,[4] or through reactive methods, could be directly fabricated as vanadium dioxide.[35,36] Wet chemistry methods utilize a precursor mixture of varying powders and acids to develop a vanadium dioxide thin film.[37,38] The doping material is introduced to the precursor film or solution with the molar ratio between vanadium and dopant atoms carefully controlled while additional annealing for further diffusion of the dopant atoms is usually required.[32] However, most of fabrication methods in fabricating high-quality tungsten-doped $VO_2$ thin films suffer from expensive operations, slow deposition rate, and highly selective substrate materials.

This study aims to develop a scalable and cost-effective process for the fabrication of tungsten-doped vanadium dioxide thin films with lowered transition temperature via co-sputtering, furnace oxidation and thermal annealing. Tungsten doped vanadium dioxide thin film samples with different doping levels up to 4.1 at.% are fabricated, and materials characterizations are carried out to confirm the tungsten doping concentrations. Parametric studies on temperature and time are performed to determine optimal conditions in furnace oxidation and thermal annealing. The insulator-to-metal phase transition behaviors of fully annealed samples are comprehensively studied with temperature-dependent infrared spectroscopy in a wide temperature range from -60°C to 100°C. The dependence of tungsten doping level on the phase transition temperature reduction and thermal hysteresis is quantitatively discussed.

## II. EXPERIMENTAL METHODS
### A. Co-Sputtering of vanadium and tungsten

Fabrication of tungsten doped $VO_2$ thin films begins with the co-sputtering of vanadium and tungsten via magnetron sputtering (Lesker PVD75) of dual vanadium and tungsten targets (Kurt J. Lesker Co., 99.95% purity), as shown in Fig. 1(a). The base pressure of the sputtering chamber was $5\times10^{-6}$ Torr and the deposition pressure was 5 mTorr. A power of 300 W was provided to the vanadium target which yields a deposition rate of 1.5 Å/s. Simultaneously a much smaller power (11 to 20 W) was provided to the tungsten target, which yields a much smaller deposition rate (0.022 to 0.1 Å/s). Note that the deposition rate of vanadium and tungsten were measured beforehand from depositing each target individually at a given power at much longer deposition time. Different tungsten doping level was achieved by varying the power to deposit tungsten, while the power to deposit vanadium was fixed at 300 W. Gun shutters were not opened until the plasma power and deposition rate became stable. The gun shutters for both targets were kept open for a total of 200 seconds, which leads to deposition of 30-nm-thick tungsten-vanadium thin films onto precleaned undoped silicon wafer substrate (280-mm-thick, double-side polished, resistivity >10,000 Ω-cm). The wafer substrate was mounted on the rotary stage which ensures the film uniformity during the co-sputtering process. The doping level or the atomic percentage of the tungsten in the co-sputtered vanadium thin film is estimated by the predetermined deposition rates for the given powers as

$$x\ (at.\%) = \frac{r_{at,W}}{r_{at,W} + r_{at,V}} \quad (1)$$

where $r_{at} = r\rho/M$ is the atomic deposition rate of tungsten or vanadium with deposition rate $r$, density $\rho$ and molar mass $M$. Multiple vanadium samples with different tungsten doping levels (including undoped) were deposited following the same co-sputtering process for the furnace oxidation and thermal annealing to fabricate tungsten doped $VO_2$ thin films.

## B. Furnace oxidation and thermal annealing

The co-sputtered tungsten-vanadium thin films were first oxidized in a tube furnace (Thermco Minibrute) with carefully controlled gas flow, temperature and time and then thermally annealed to fully diffuse tungsten atoms. Following our previous successful furnace oxidation process,[4,5] the $N_2$ and $O_2$ gas flowrates were respectively set to 7 SLPM and 0.5 SLPM, while the oxidation temperature was fixed at 300°C. The oxidation time is critical, as the $VO_2$ thin film could be under-oxidized with inadequate time, or over-oxidized with excessive time. For 30-nm-thick vanadium thin film deposited on silicon substrate by e-beam evaporation, it requires 3 hours to fully oxidize the film into $VO_2$. However, sputtered vanadium film of the same thickness could have different grain size, porosity and surface roughness other than that prepared by e-beam evaporation. Therefore, an oxidation time study needs to be conducted for determining the optimal oxidation time for the sputtered vanadium thin films. The thermal annealing was performed in the same tube furnace with only $N_2$ gas flowing at 7 SLPM and $O_2$ gas flow shut off to avoid over-oxidizing the $VO_2$ film to other vanadium oxides like $V_2O_5$. A comprehensive thermal annealing study on the temperature and time was conducted to determine the optimal condition to fully anneal the tungsten doped $VO_2$ thin films at different doping levels.

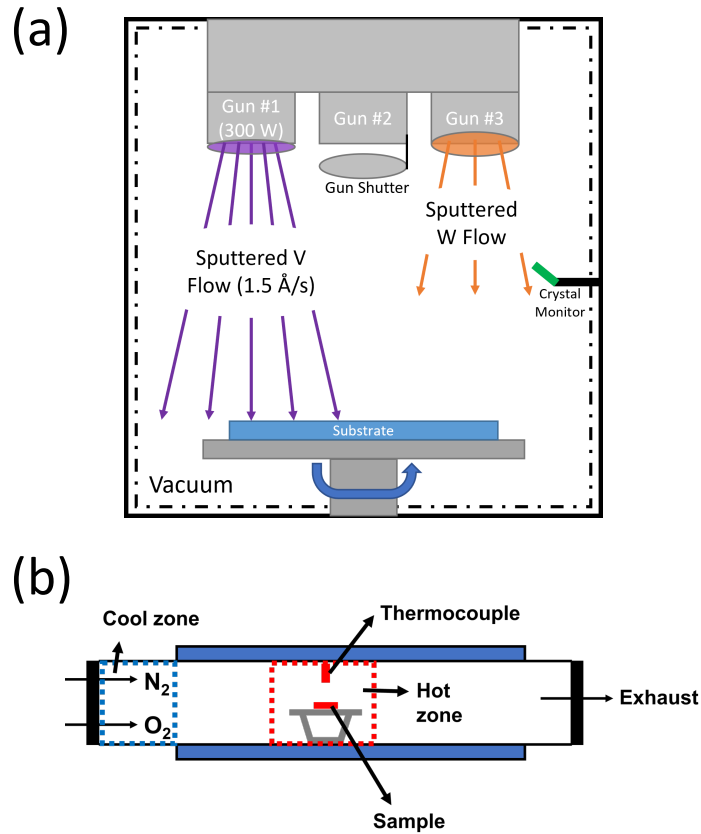

Fig. 1. Schematics for fabricating tungsten-doped vanadium dioxide thin films ($W_xV_{1-x}O_2$) with lowered phase transition temperatures via (a) magnetron co-sputtering of vanadium and tungsten and (b) furnace thermal oxidation and thermal annealing.

C. Temperature dependent infrared transmittance measurement

In order to study the insulator-to-metal phase transition behaviors of undoped and doped $VO_2$ thin films on undoped silicon wafer substrate, temperature dependent infrared transmittance was characterized using a Fourier-transform infrared spectrometer (Thermo Fisher, Nicolet iS50 FTIR) in the spectral range from 2 to 25 μm in wavelength at a resolution of 8 cm$^{-1}$ with each spectrum averaged over 32 scans. To facilitate the oxidation and thermal annealing studies, a home-built ambient heating stage with an 8-mm clear aperture, whose temperature can be precisely controlled from 25°C to 120°C with a PID temperature controller and polyimide thin-film heater, was used to mount the sample for the temperature-dependent FTIR transmittance measurement in ambient. To characterize fully annealed tungsten doped $VO_2$ thin films whose phase transition could go below room temperature, an optical cryostat (Janis VPF-800-FTIR) custom-built to be mounted inside the FTIR sample compartment with optical pass through ZnSe windows was used to vary the sample temperatures in a wider range from -60°C to 100°C. At each sample temperature, 5 minutes were given to ensure the steady state was achieved before the spectrum was taken. The temperature dependent transmittance of $WVO_2$ thin film were measured by 5°C intervals in both heating and cooling processes to study the thermal hysteresis behavior during the phase transition.

**D. XPS characterization**

X-ray photoelectron spectroscopy (XPS, Kratos Axis Supra[+]) was conducted with and Al k$\alpha$ 1486.6 eV X-ray source in ultrahigh vacuum. V2p elemental scans from binding energies 510.0 eV to 540.0 eV were performed to identify the oxidation state of vanadium, and W4f elemental scans from binding energies 30.0 eV to 50.0 eV were carried out to measure the atomic concentration of tungsten in the fully annealed $W_xV_{1-x}O_2$ samples. Raw data was analyzed and fitted with CasaXPS software, while Shirley algorithm was utilized to draw background for both V2p and W4f elemental scans. Lorentzian Asymmetric curve was used to fit the V2p, V3p and O1s peaks while Gaussian Lorentzian peaks were used to fit the peaks of W4f.

## III. RESULTS AND DISCUSSION

**A. Oxidation time study of $VO_2$ thin films**

It is not trivial to obtain high-quality $VO_2$ thin films by oxidizing vanadium thin films. In our previous work,[4,5] optimal oxidation conditions such as temperature, gas flowrates and time have been identified for fully oxidizing 30-nm vanadium thin film on silicon wafers prepared from e-beam evaporation. Among them, the film quality is sensitive to the oxidation time as it could be under or over oxidized with insufficient or excessive time. As the vanadium and doped vanadium thin films were prepared by sputtering process, it is necessary to check the oxidation time again to achieve full oxidization of vanadium films into stoichiometric $VO_2$. To this end, 30-nm vanadium films were deposited onto undoped silicon wafers respectively with e-beam evaporation and sputtering, which were oxidized at 300°C in the furnace for different time periods.

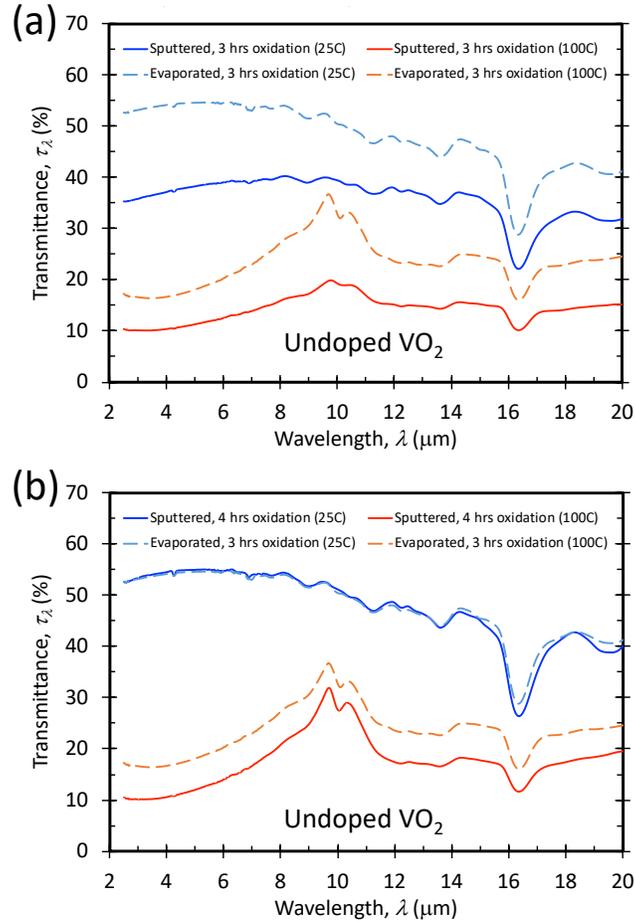

Fig. 2. Spectral infrared transmittance of undoped VO$_2$ thin films at 25°C and 100°C (a) after 3 hours thermal oxidation and (b) after 4 hours thermal oxidation in an oxygen-rich furnace at 300°C from 30-nm sputtered vanadium film in comparison with that of thermal evaporated 30-nm vanadium film fully oxidized after 3 hours.

As shown in Fig. 2(a), after 3 hours furnace oxidation, the evaporated vanadium film exhibits full oxidation state of VO$_2$ with maximum infrared transmission around 55% at 25°C in its insulating phase and lowest infrared transmission around 15% at 100°C in its metallic phase, which is consistent with our previous work[4]. However, the sputtered 30-nm vanadium thin film shows much less infrared transmission particularly about 38% at 25°C in its insulating phase. This clearly indicates that the vanadium film is not completely oxidized, and more oxidation time is required. With 4 hours oxidation, as shown in Fig. 2(b), the sputtered vanadium thin film achieves the same infrared transmission around 55% at 25°C as expected for the $VO_2$ in insulating phase. Moreover, the oxidized film also exhibits similar infrared transmission spectrum at 100°C for the $VO_2$ in metallic phase as the fully oxidized vanadium from e-beam evaporation, regardless of 3% to 5% difference which might be due to differences in grain sizes and porosity from the different deposition methods of vanadium. Note that the high transmission peaks around 10 μm wavelength observed in the metallic phase stems from the over oxidation at the film surface due to high oxygen concentration. It is reasonably assumed that slight doping of tungsten (less than 5 at.%) would not significantly affect the oxidation time. Therefore, in the following studies all the 30-nm doped vanadium films were oxidized at 300°C for 4 hours into $WVO_2$ thin films.

### B. Thermal annealing of $WVO_2$ films

In order to ensure that tungsten atoms are completely diffused and bonded with $VO_2$ to lower its IMT temperature, thermal annealing was carried out for fully oxidized $WVO_2$ thin films. Comprehensive studies were performed on the thermal annealing temperature from 400°C to 650°C and annealing time from 1 hour to 2 hours. Infrared transmittance above room temperature upon heating (25°C up to 100°C) and cooling (100°C down to 25°C) was measured with the FTIR spectrometer along with the ambient heating stage after each $WVO_2$ film was thermally annealed.

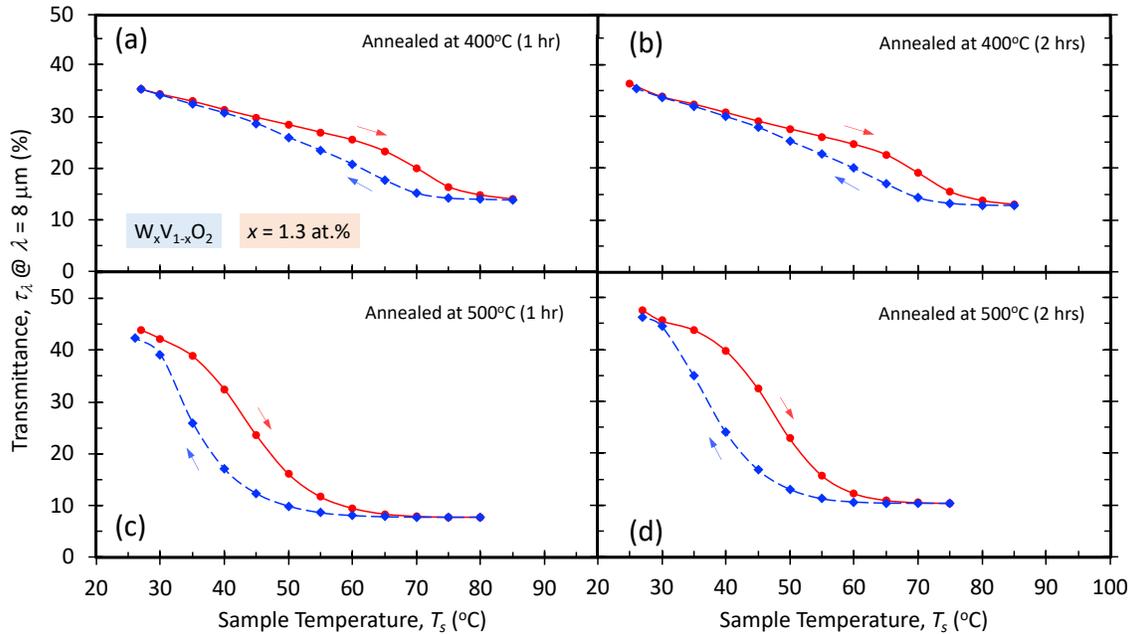

Fig. 3. Thermal annealing study with temperature-dependent transmittance at $\lambda$ = 8 μm wavelength from heating (red) and cooling (blue) between 20°C and 100°C for $W_{0.013}V_{0.097}O_2$ thin film: (a) 400°C for 1 hour, (b) 400°C for 2 hours, (c) 500°C for 1 hour, and (d) 500°C for 2 hours.

As shown in Fig. 3, the spectral transmittance at $\lambda = 8$ μm wavelength at various sample temperature was selected to illustrate the phase transition behavior upon heating and cooling for the $W_{0.013}V_{0.097}O_2$ thin film after different annealing conditions. With 400°C and 1 hour annealing, the WVO2 film with 1.3 at.% exhibits gradual IMT transition mainly occurring between 40°C and 80°C. However, when the annealing temperature increases to 500°C, the IMT transition becomes abrupt with lowered temperature ranges from 25°C and 65°C. No noticeable changes were observed from thermal annealing with an additional hour, which suggests one hour is sufficient for thermally annealing the fabricated WVO2 thin films.

To get a better idea on the thermal annealing effect on the IMT of the $W_{0.013}V_{0.097}O_2$ thin film, the derivative of the infrared transmittance at $\lambda = 8$ μm wavelength with respect to sample temperature, i.e., $\Delta\tau_\lambda/\Delta T_s$, is plotted in Fig. 4. The $W_{0.013}V_{0.097}O_2$ thin film changes its infrared transmittance at most by −0.75%/K after 400°C annealing, which increases up to −1.75%/K upon heating or even −2.65%/K upon cooling after 500°C annealing. The corresponding temperature for the largest rate upon heating drops from 72°C to 42°C for the thermal annealing from 400°C to 500°C. This confirms that thermal annealing at 500°C is sufficient for the $W_{0.013}V_{0.097}O_2$ thin film to effectively lower its IMT temperature by 30°C.

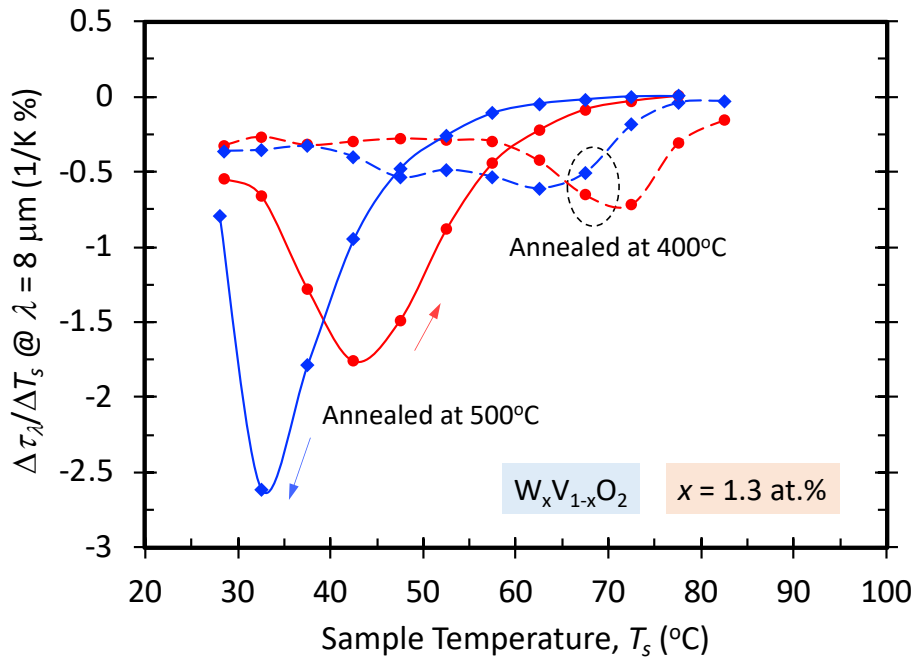

Fig. 4. Transmittance derivative with respect to temperature at $\lambda = 8$ μm wavelength from heating (red) and cooling (blue) between 20°C and 100°C for $W_{0.013}V_{0.087}O_2$ thin film thermally annealed under 400°C (dash) and 500°C (solid) for 1 hour.

Figure 5 shows the spectral transmittance at $\lambda = 8$ μm wavelength at temperatures from 25°C to 100°C for the $W_{0.021}V_{0.079}O_2$ thin film thermally annealed at 500°C, 550°C, 600°C, and 650°C after 1 hour, where apparent changes in transmittance with higher annealing temperature are observed. The phase transition upon heating and cooling shifts to lower sample temperatures with the film annealed at higher temperature. In particular, the difference between 600°C and 650°C becomes marginal, indicating that the thermal annealing is effective at these higher temperatures. Therefore, for the $W_{1-x}V_xO_2$ films with tungsten doping levels greater than 2 at.%, thermal annealing at 650°C is adopted to effectively lower IMT temperatures.

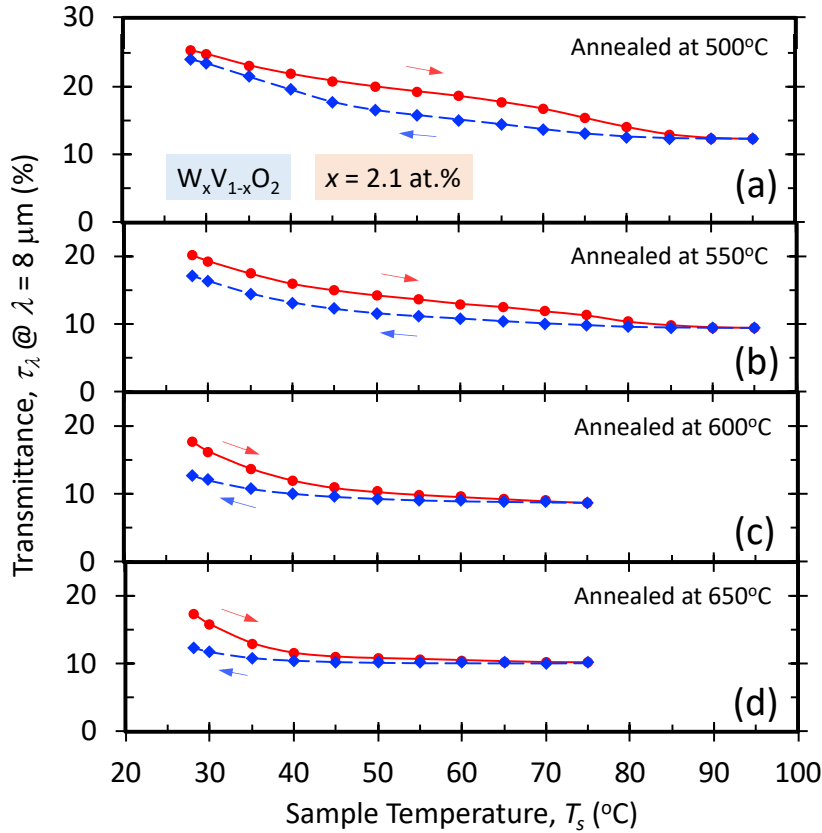

Fig. 5. Thermal annealing study with temperature-dependent transmittance at $\lambda = 8$ μm wavelength from heating (red) and cooling (blue) for $W_{0.021}V_{0.079}O_2$ thin film: (a) 500°C, (b) 550°C, (c) 600°C, and (d) 650°C for 1 hour.

Note that the spectral transmittance at 25°C is only 18% upon heating and 12% upon cooling, while the difference between heating and cooling indicates that the phase transition could continue at sub-ambient temperatures. This is more clearly observed with the rate of transmittance change as $\Delta\tau_\lambda/\Delta T_s$ shown in Figs. 6(a) and 6(b) respectively upon heating and cooling for the $W_{0.021}V_{0.079}O_2$ thin film thermally annealed at different temperatures. The rate becomes flat at temperature above 50°C with a large drop around room temperature after 650°C annealing, indicating the lowered IMT transition temperatures towards room temperature and sub-ambient IMT transition. The $W_{0.021}V_{0.079}O_2$ film has similar behaviors upon cooling but the rate is smaller compared to the heating due to thermal hysteresis. The observation with sub-ambient IMT phase transition with the $W_{0.021}V_{0.079}O_2$ film is expected because the IMT temperature of $VO_2$ could be lowered by 20°C to 30°C per one at.% tungsten doping as reported by other fabrication methods.[30,33,34]

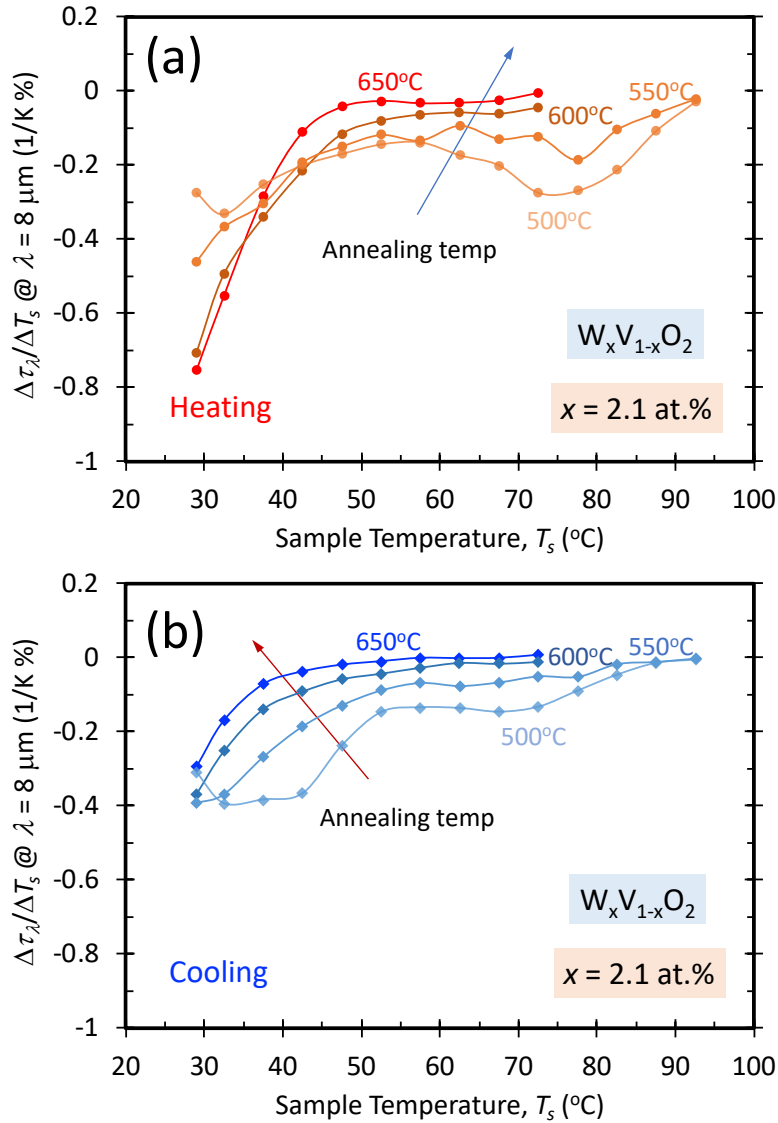

Fig. 6. Transmittance derivative with respect to temperature at $\lambda$ = 8 μm wavelength from (a) heating and (b) cooling for $W_{0.021}V_{0.079}O_2$ thin film thermally annealed under 500°C, 550°C, 600°C, and 650°C for 1 hour.

## C. IMT characteristics of fully annealed WVO$_2$ films

In order to fully characterize the IMT behaviors of W$_{1-x}$V$_x$O$_2$ films, particularly with tungsten doping level ($x$) greater than 2 at.%, whose phase transition could go to sub-ambient temperatures, an FTIR cryostat was used to measure the infrared transmittance at a much wider temperature range from -60°C to 100°C. Figure 7 shows the spectral infrared transmittance at steady sample temperatures upon heating and cooling with 5°C intervals for undoped VO$_2$ (not annealed) and several fully annealed W$_{1-x}$V$_x$O$_2$ films with tungsten doping levels of 1.3 at.%, 2.6 at.%, 3.3 at.%, 3.7 at.%, and 4.1 at.%, fabricated from the aforementioned co-sputtering, furnace oxidation and thermal annealing processes. It can be observed that, the transmittance in the insulating phase decreases from ~60% to ~40% with tungsten doping from 0 to 4.1 at.%. In the metallic phase, the annealed WVO$_2$ thin films exhibit much flatter infrared transmission spectra around 10% compared to that of undoped VO$_2$ without annealing. It is believed that this is due to the much lower oxygen trace in the furnace during the annealing process with only N$_2$ gas flow, which reduces most of the surface over-oxides (i.e. V$_2$O$_5$) to VO$_2$.

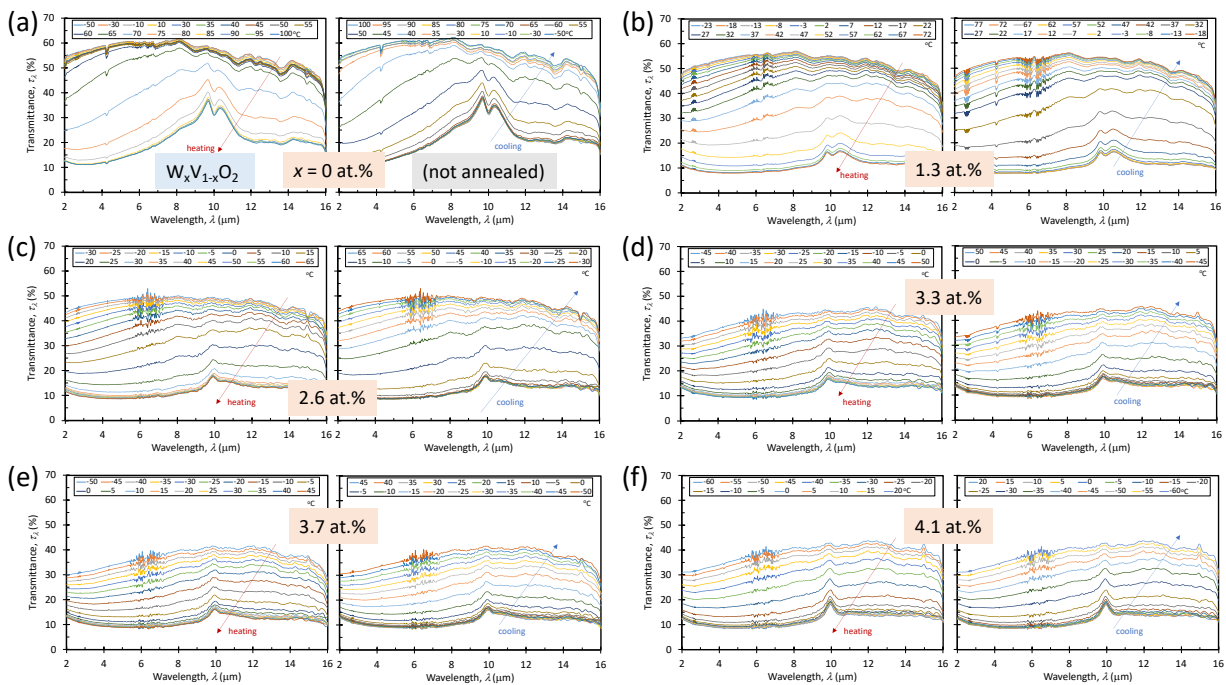

Fig. 7. Spectral infrared transmittance between −60°C and 100°C from heating and cooling of (a) undoped VO2 thin film (not annealed) and tungsten doped W$_x$V$_{1-x}$O$_2$ thin films (fully annealed) with different tungsten doping level: (b) 1.3 at.%, (c) 2.6 at.%, (d) 3.3 at.%, (e) 3.7 at.%, and (f) 4.1 at.%, fabricated from co-sputtering and furnace thermal processes.

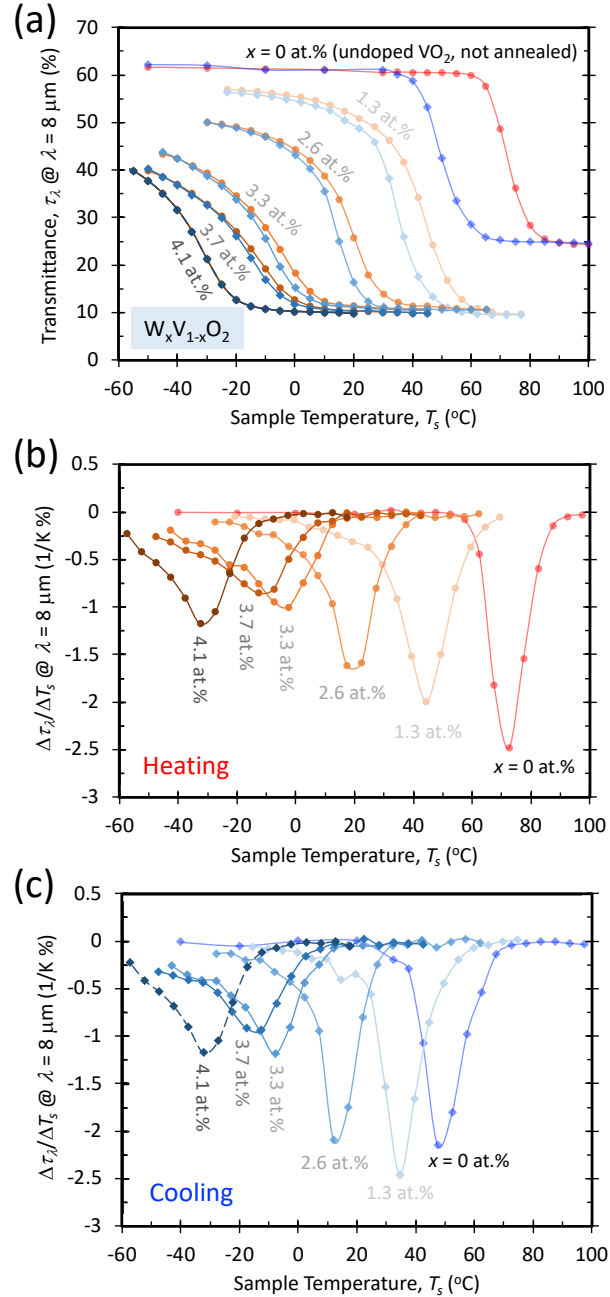

Fig. 8. Insulator-to-metal transition characteristics between -60°C and 100°C of $W_xV_{1-x}O_2$ thin films with different tungsten doping levels from x=0 to 4.1 at.% fabricated from co-sputtering and furnace thermal processes: (a) temperature-dependent transmittance at $\lambda = 8$ μm wavelength from both heating (red) and cooling (blue) and transmittance derivative with respect to temperature from (b) heating and (c) cooling individually.

To better understand the IMT behaviors of fully annealed $WVO_2$ films, spectral transmittance at $\lambda = 8$ μm is plotted as a function of sample temperature upon both heating and cooling processes for all the samples in Fig. 8(a). Clearly, with more tungsten doping, the insulator-to-metal phase transition successfully shifts to lower temperatures along with the thermal hysteresis between heating and cooling becoming smaller. Figures 8(b) and 8(c) present the derivative of the infrared transmittance at $\lambda = 8$ μm wavelength with respect to sample temperature, i.e., $\Delta \tau_\lambda/\Delta T_s$, for heating and cooling processes, respectively. The IMT derivative valley moves to the lower temperatures along with smaller magnitude with higher tungsten doping. In particular, the IMT midpoint temperatures upon heating, $T_{m,heat}$, at which there exists the largest derivative magnitude, are 72.5, 44.5, 20, -2.5, -12.5, and -32.5°C for unannealed $VO_2$ and fully annealed $W_{1-x}V_xO_2$ films with doping levels of $x = 0$, 1.3, 2.6, 3.3, 3.7, and 4.1 at.%.

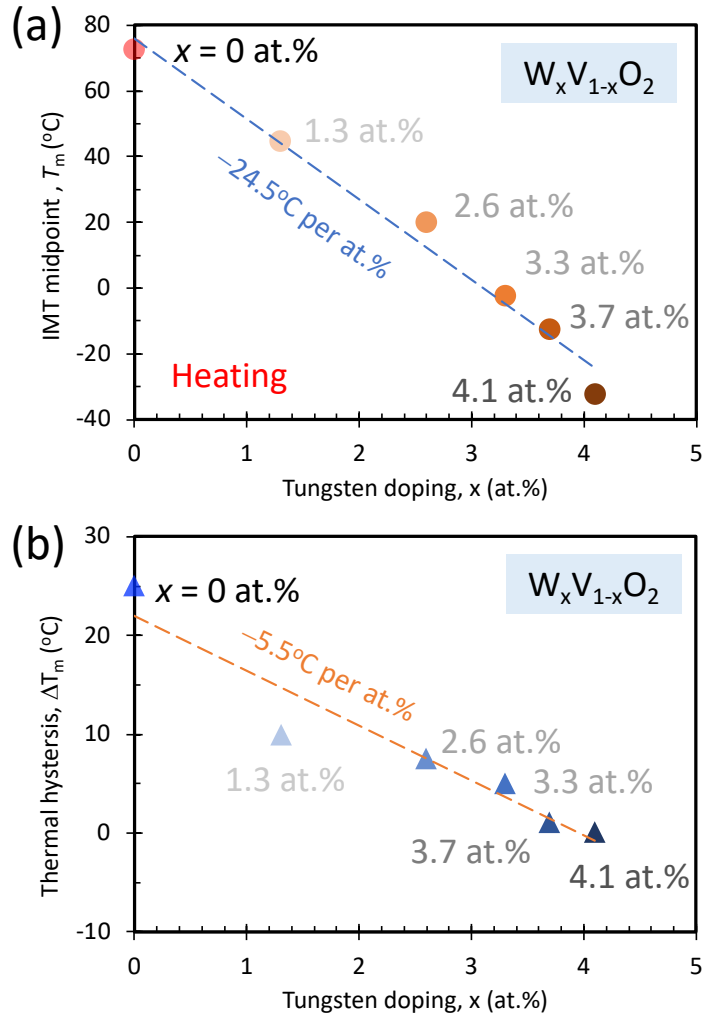

Fig. 9. (a) Insulator-to-metal transition midpoint from heating and (b) thermal hysteresis between heating and cooling of $W_xV_{1-x}O_2$ thin films as a function of tungsten doping levels from x=0 to 4.1 at.% fabricated from co-sputtering and furnace thermal processes.

Figure 9(a) reveals the dependence of the IMT midpoint temperature upon heating, $T_{m,heat}$, on the tungsten doping level as a linear correlation with a slope of –24.5°C per at.% in lowering the IMT temperature with tungsten doping. Thermal hysteresis behavior during IMT is also known for VO$_2$ thin films. By taking the difference in the IMT midpoint temperature between the heating and cooling processes, i.e., $\Delta T_m = T_{m,heat} - T_{m,cool}$, the thermal hysteresis is 25, 10, 5, 1, and 0°C for unannealed VO$_2$ and fully annealed W$_{1-x}$V$_x$O$_2$ films with doping levels of $x$ = 0, 1.3, 2.6, 3.3, 3.7, and 4.1 at.%. As shown in Fig. 9(b), a linear fitting suggests the thermal hysteresis decreases with tungsten doping level at a slope of –5.5°C per at.% from the W$_{1-x}$V$_x$O$_2$ films with $x$ from 0 to 4.1 at.% fabricated from the co-sputtering, furnace oxidation and thermal annealing processes.

**D. Tungsten doping level from XPS depth profiling**

Finally, the tungsten doping level or the atomic concentration was confirmed with XPS characterization for the fully annealed W$_x$V$_{1-x}$O$_2$ sample with $x$ = 3.7 at.% estimated from the sputtering deposition rates of tungsten and vanadium targets according to Eq. (1). Figure 10(a) shows the XPS spectra as function of binding energy at the film surface, where high W4f$_{7/2}$ and W4f$_{5/2}$ peaks are observed at 34.9 eV and 37.1 eV, confirming the successful thermal annealing. However, the measured atomic concentration of tungsten at the film surface is 14.8 at.%, which is much higher than the 3.7 at.% from the estimation. Note that this cannot represent the tungsten concentration within the film, as it could be affected by surface termination conditions, and XPS characterization within the film was required for more accurate measurement of tungsten concentration. Therefore, about 25 nm of the W$_{0.037}$V$_{0.063}$O$_2$ film was etched away with 5 keV Ar$^+$ ions for 10 mins to reach the center of the film. As shown in Fig. 10(b), the W4f peaks become much lower. With another 10 mins of etching, Fig. 10(c) presents the XPS spectrum at the bottom of the W$_{0.037}$V$_{0.063}$O$_2$ film, which is almost the same as that at the center of the film, indicating excellent uniformity within the film from the co-sputtering and thermal furnace processes. The measured tungsten concentration was 3.88 at.% at the center and 3.62 at.% at the bottom of W$_{0.037}$V$_{0.063}$O$_2$ film, or averaged to be 3.75±0.18 at.%, which is consistent with the estimated value of 3.7 at.%. This confirms the accurate estimation on the tungsten doping levels from the sputtering deposition rates.

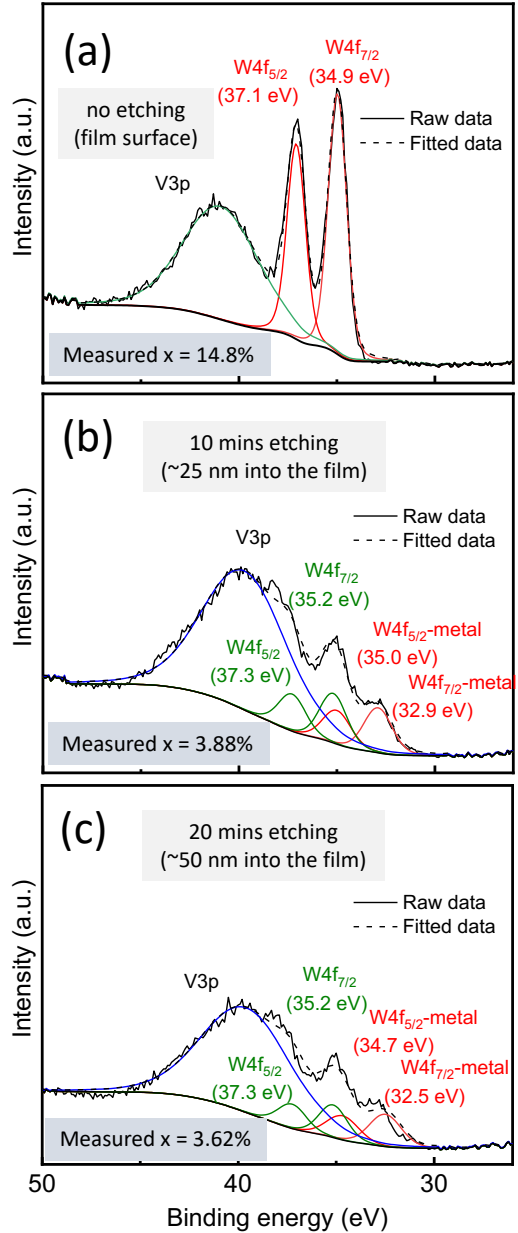

Fig. 10. XPS spectra of the $W_{0.037}V_{0.063}O_2$ thin film (a) at the surface, (b) at ~25 nm into the film after 10 mins etching and (c) at ~50 nm into the film after 20 mins etching with Ar+ ions.

## IV. CONCLUSION

In summary, we have successfully demonstrated that the phase transition temperature is lowered with tungsten doped vanadium dioxide thin films fabricated via co-sputtering, furnace oxidation and thermal annealing processes. By varying the sputtering power for the tungsten target, vanadium thin films were doped with tungsten at different concentrations. For 30-nm sputtered vanadium thin films on undoped silicon wafer, optimal oxidation time was found to be 4 hours in an oxygen-rich furnace at 300°C. Also, lightly tungsten-doped vanadium dioxide thin films with doping level of 1.3 at.% could be fully annealed at 500°C, while higher annealing temperature up to 650°C was required to achieve full annealing for heavily doped ones with atomic concentration greater than 2 at.%. Comprehensive temperature-dependent infrared transmittance measurements of fully annealed tungsten-doped vanadium dioxide thin films up to 4.1 at.% revealed the phase transition temperature decreases at 24.5°C per at.% of tungsten doping along with 5.5°C per at.% shrinkage of thermal hysteresis between heating and cooling. The results in this study would facilitate the research and development of thermochromic coatings for wide energy and thermal control applications.

## ACKNOWLEDGMENTS

This work was support in part by National Science Foundation (CBET-2212342) and National Aeronautics and Space Administration (80NSSC21K1577).

## AUTHOR DECLARATIONS

### Conflict of Interest

The authors have no conflicts to disclose.

### Author Contributions

V.K.R. and J.C. fabricated the samples, measured the FTIR transmittance with ambient heating stage; J.C. did the thermal annealing at temperatures up to 500°C; V.K.R. did the thermal annealing at temperatures above 500°C; V.K.R. conducted XPS characterizations; L.W. performed the transmittance measurement with FTIR cryostat for all fully annealed samples; L.W. and S.T. conceived the idea, secured funds and supervised the project; L.W., V.K.R., and J.C. prepared the figures and wrote the manuscript; All authors reviewed the manuscript.

## DATA AVAILABILITY

The data that support the findings of this study are available from the corresponding author upon reasonable request.